\documentclass[a4paper,11pt]{article}
\usepackage{pos}
\usepackage{subcaption}
\usepackage{gensymb}
\usepackage{siunitx}
\usepackage{wrapfig}

\newcommand{\refer}[1]{Ref.~\cite{#1}}

\title{Cosmic Ray Detection with the IceTop Enhancement}

\author{Megha Venugopal$^{a,*}$ on behalf of the IceCube Collaboration}
\renewcommand{\printHeadAuthors}{Megha Venugopal}

\affiliation[a]{Karlsruhe Institute of Technology, Institute for Astroparticle Physics, 76021 Karlsruhe,
Germany\\}
\affiliation[*]{$\mathrm{Speaker}$}
\emailAdd{megha.venugopal@kit.edu}

\abstract{ IceTop is the cosmic-ray detector located on the surface of the IceCube Neutrino Observatory at the South Pole, consisting of 81 pairs of ice-Cherenkov tanks. The rise in the energy threshold of air-shower measurements in IceTop due to accumulating snow emphasized the need for the next generation of IceCube surface detectors. For this purpose, the Surface Array Enhancement (SAE) is set to comprise elevated scintillator panels and radio antennas controlled by hybrid DAQ systems. The detectors of the SAE are also expected to extend to the planned IceCube-Gen2 Surface Array. An initial study with a prototype station is already conducted. We briefly review the SAE and the deployment as well as the calibration status of the upcoming stations of the planned array of 32 stations. The focus of this contribution is on the radio detection of extensive air showers. A preliminary estimation of the position of the shower maximum ($X_\mathrm{max}$), that is sensitive to the primary mass, with data from the 3 antennas of the prototype station was carried out. An extension of the method from previous analyses is also briefly discussed.}

\FullConference{The European Physical Society Conference on High Energy Physics (EPS-HEP2023)\\
 21-25 August 2023\\
Hamburg, Germany\\}

\begin{document}
\maketitle


\section{Introduction} \label{1}
The IceCube Neutrino Observatory at the geographic South Pole consists of a cubic kilometer of instrumented ice measuring the Cherenkov light from neutrino-induced lepton showers. IceTop is the surface cosmic ray detector that has contributed significantly to the knowledge of the cosmic-ray spectrum and its composition at PeV energies.

The Surface Array Enhancement is a planned 32 station setup, each equipped with 3 antennas and 8 scintillators, all elevated with extendable legs, in order to increase threshold and mitigate rising snow accumulation on IceTop tanks. 
The scintillators observe the charge deposit with respect to a minimum ionizing particle (MIP). The antennas measure the electromagnetic emission by the air-shower particles in the geomagnetic field and due to the Askaryan effect \cite{FrankRadioPaper}. IceTop measures the Cherenkov light emitted by charged particles from shower cascades in its ice-Cherenkov tanks. Cosmic-ray observation at the Surface Array Enhancement will combine the data from scintillators, antennas, IceTop and in-ice detectors. The combination of inputs from all detection modes will improve our understanding of air-shower and cosmic-ray physics in the PeV - EeV primary energy range.
 

\section{Calibration and production status of the Surface Array Enhancement} \label{2}

\begin{figure}[h]

\begin{subfigure}{0.5\textwidth}
\includegraphics[width=0.8\linewidth, height=6cm]{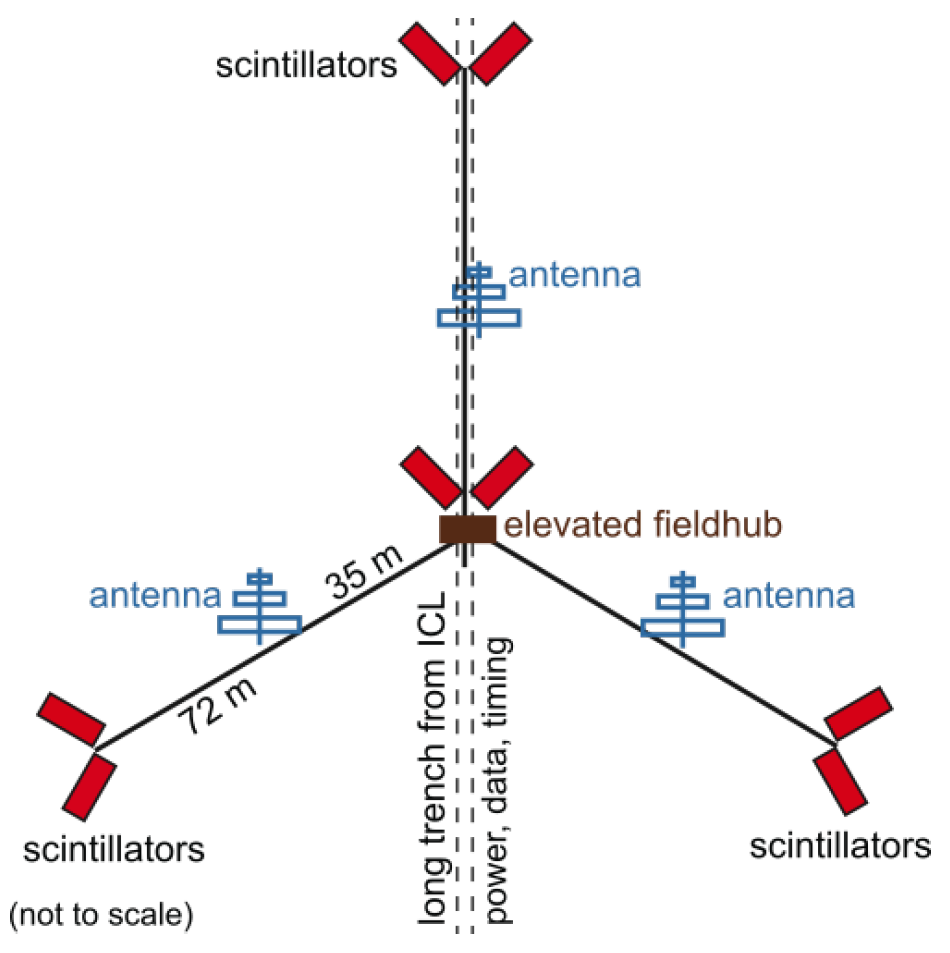} 
\caption{The layout of a single station of the SAE}
\label{fig:EnhancementLayout}
\end{subfigure}
\begin{subfigure}{0.5\textwidth}
\includegraphics[width=0.7\linewidth, height=7cm]{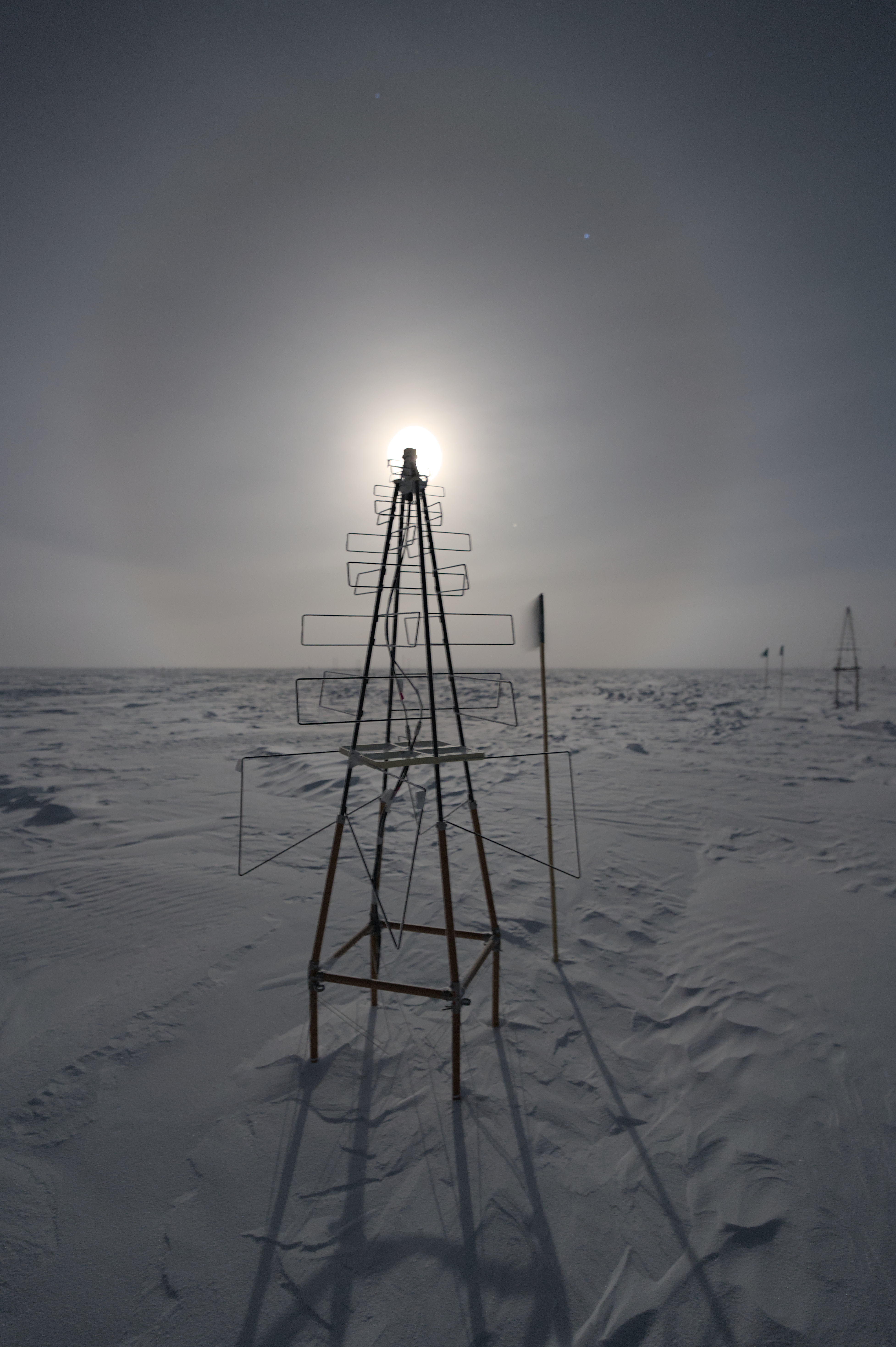}
\caption{The Log Periodic Dipole Array (LPDA) Antenna of the Enhancement}
\label{fig:Antenna}
\end{subfigure}

\caption{Left: A single station with 8 scintillators, 3 antennas and the central fieldhub containing the TAXI inside. Right: The antenna of the SAE with the production mount. The top of the antenna contains the Low Noise Amplifier and its housing. Picture credit: Marc Jacquart.}
\label{fig:image2}
\end{figure}

A prototype station was deployed at the South Pole in January 2020. The configuration of the setup of a single station was based on the design shown in Fig. \ref{fig:EnhancementLayout} which is extended to 32 stations around IceTop. Several efforts were made to calibrate the central data acquisition (DAQ) referred to as the TAXI. More details on the initial setup can be found in \refer{TurcotteOehlerICRC21}. 

In January 2023, all 8 scintillators of the prototype station and one antenna were replaced in order to pave way for the first station, fully deployed with the intended production design of the Enhancement, dubbed Station 0.

For the case of deploying future stations, measurements were taken in order to characterize the radio data. A sampling rate of 1\,GHz is currently used and efforts into understanding data being sampled at lower frequencies is underway. More details on antenna design and hardware can be found in \refer{SKALAV2}. The LPDA antennas that were used can potentially capture all frequencies in the band from 50 -- 650\,MHz along two horizontal polarizations which are treated independently both during data taking and analysis. As the radio data along each polarization travels through the signal chain, it gets amplified, split into 4 channels and filtered by band-pass filters in the frequency range from 50 -- 350\,MHz \cite{thesisRox}. Amplitude calibration is carried out by passing a sine wave of known amplitude sweeping through frequencies from 50 -- 450\,MHz across the TAXI board and measuring the response of the electronics. 
\begin{wrapfigure}{l}{0.5\textwidth}
\begin{center}
\includegraphics[scale=0.75]{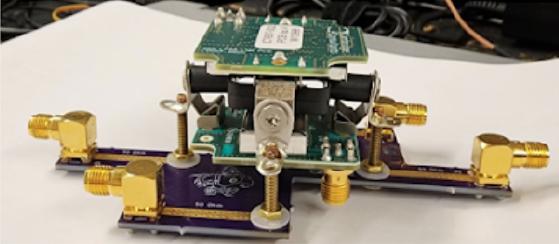}
\end{center}
\caption{The setup used to calibrate the TOP and BOT LNA boards in the anechoic chamber with a unique calibration board. The inclusion of the LNA spacers and screws is to make it as close as possible to the working setup. Figure taken from \refer{IsabellaCalibration}.}
\label{fig:LNACalibration}
\end{wrapfigure}
The Low Noise Amplifier (LNA) is a differential board consisting of the TOP and the BOT boards. The naming is based on its placement on the top and bottom inside the LNA housing of the antenna as seen in Figure \ref{fig:Antenna}. The calibration of the LNA was done in an anechoic chamber located at the University of Kansas. More details on the procedure can be found in \refer{IsabellaCalibration}. An interesting observation was the sensitivity of the TOP and BOT LNA boards to its orientation with respect to each other. A calibration board was set up to accommodate both LNA boards in the same configuration as they would be housed in the antenna. The calibration board and the mounted LNA are shown in Figure \ref{fig:LNACalibration}. The gain for both boards were characterized individually and optimal pairs with the closest characteristics were noted.

The scintillators were also extensively tested. This involved functionality and calibration runs with both the operational modes provided for the scintillators' electronics, coincidence measurements per station, as well as threshold, and voltage scans. These measurements were performed at varied temperatures, which was done to implement the firmware automation in the coming seasons, with which we aim to achieve a fixed gain and therefore, trigger rate, for all scintillation detectors. The details and further information on the scintillators of Station 0 are presented in \refer{ShefaliScintICRC}.

\section{Identification of air showers in Radio} \label{3}

Air-shower identification in the single station of the SAE uses different criteria amongst the different detectors. A hit in a scintillator is characterized by a charge deposit greater than a pre-defined threshold \cite{ShefaliScintICRC}. If more than 3 scintillator panels receive a hit, within a 1\,$\mu$s window, then a scintillator event is included into the processing. The hardware trigger condition for antenna readout requires a hit in 6 or more scintillator panels within a 1\,$\mu$s window. The trigger condition is easily modifiable as necessary when extending to more stations. Data from IceTop undergoes processing at earlier stages of software filtering and is stored with information on the energy proxy $S_\mathrm{125}$ and directional reconstruction - zenith and azimuth using the standard reconstruction \cite{S125IceTop}. In both the below mentioned methods, a lack of radio pulses with a zenith below 20\degree\ is observed due to the uniqueness of the detector location at the South Pole. The low geomagnetic angle at smaller zenith angles leads to a lesser geomagnetic effect contribution to the overall radio pulse when compared to other locations \cite{NarayanAskaryan}.

\subsection{Conventional Approach}

\begin{figure}[h]
    \centering
    \includegraphics[width=0.7\linewidth]{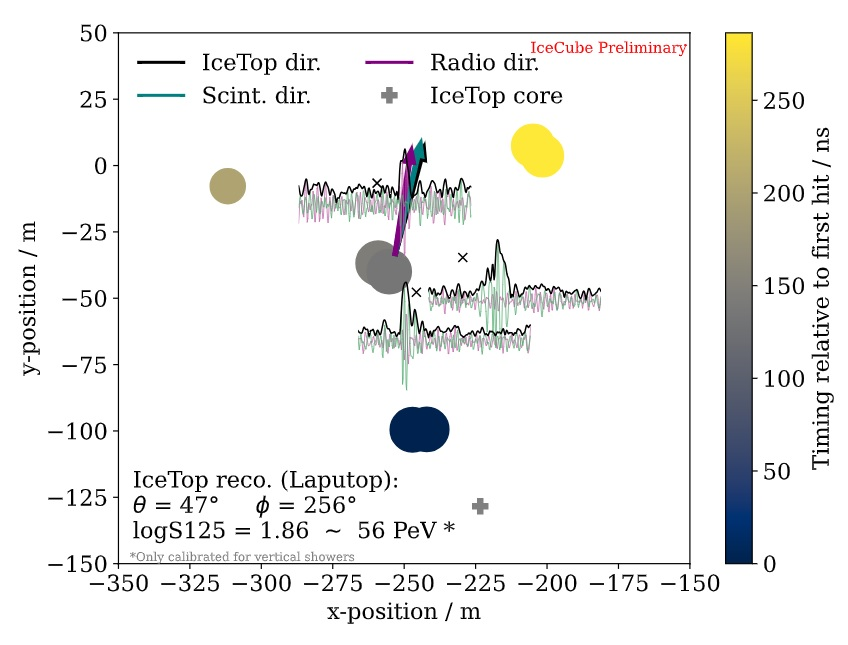}
    \caption{Figure indicates an example air shower identified across different detectors using the method from \refer{HrvojeShowerIdentification}. The size of the circles represent the charge deposit in the scintillator and the colour indicates time. The black crosses indicate the antenna positions.}    \label{fig:IdentifiedAirShower}
\end{figure}

Air-showers are identified by a triple coincidence method. This uses data from radio, scintillator and IceTop and checks if the event timestamps of all of them are within a 2\,$\mu$s window. This technique was first applied in \refer{HrvojeShowerIdentification} and subsequently extended to measured data up to January 2022. 
For the radio data, the median of the recorded data is taken over multiple traces in order to reduce the effect of spikes and a background frequency filter is applied to reduce frequency specific Radio Frequency Interference (RFI).
It is possible that a hardware trigger is sent upon coincidence with scintillators but there is no identifiable radio pulse in the traces. This is primarily due to the differences in the energy threshold for the two detectors. In order to find air-shower radio pulses, multiple quality cuts are applied which included a Signal to Noise Ratio (SNR) high enough to distinguish the pulse from noise as well as a check for the possible temporal pulse location with respect to actual antenna positions. Figure \ref{fig:IdentifiedAirShower} shows an air shower identified with the conventional method.

\subsection{The Convolutional Neural Network (CNN) Approach}

 A second approach uses CNN to identify radio pulses in high background \cite{AbdulCNN}. A \textit{Classifier} was used for identifying signal from background and a \textit{Denoiser} was used to clean the traces from the background RFI. 

 \begin{figure}[h]
     \centering
     \includegraphics[width=0.8\linewidth]{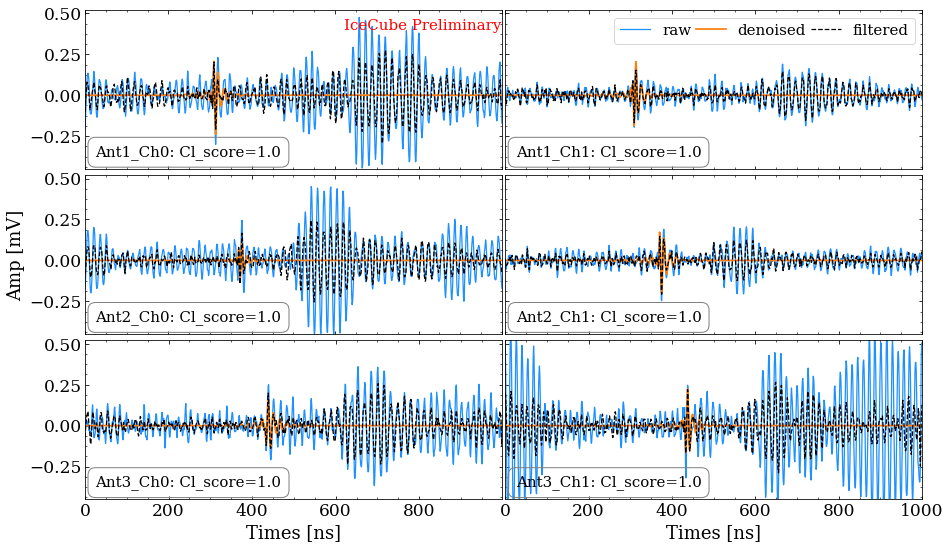}
     \caption{Figure from \refer{AbdulCNN} where the raw waveforms are represented in blue, waveforms after using the \textit{Denoiser} in orange and after the use of a spike filter with black dashed lines.}
     \label{fig:CNNradioidentification}
 \end{figure}
CoREAS \cite{CoREAS} simulations were generated for the single station which produced radio pulses corresponding to each antenna. Noise from soft triggered data was averaged over several traces and the pulse was re-injected to mimic measured data. This was used to train the network. The pulse identification was then carried out with the measured data.
Figure \ref{fig:CNNradioidentification} shows the raw waveform and the waveform extracted after the use of the \textit{Denoiser} network. The trace is compared with the use of a spike filter which is also described as the inverted spectrum filter in \refer{Xmaxpaper}. This work was carried out with the processed data from January and February 2022.

\section{Estimation of $X_\mathrm{max}$ with the identified air-showers} \label{4}

The depth of shower maximum, $X_\mathrm{max}$, directly provides an insight into the mass of the primary particle. The method used for the estimation of $X_\mathrm{max}$ has already been extensively studied in the LOFAR experiment \cite{XmaxLOFAR}. An in-depth analysis of implementing this method to the prototype station of the IceTop Enhancement can be found in \refer{thesisRox}. The work was extended to all the measured air showers of the Enhancement in \refer{Xmaxpaper} obtained using the two approaches described in Section~\ref{3}. It involves running multiple simulations to estimate $X_\mathrm{max}$ for each reconstructed energy estimate and other parameters from IceTop. The simulations were run for proton and iron primaries with CORSIKA \cite{CORSIKA}, an air-shower simulation program which, using CoREAS, is capable of generating radio pulses corresponding to the air shower. 

\begin{figure}
    \centering
    \includegraphics[width=0.8\linewidth]{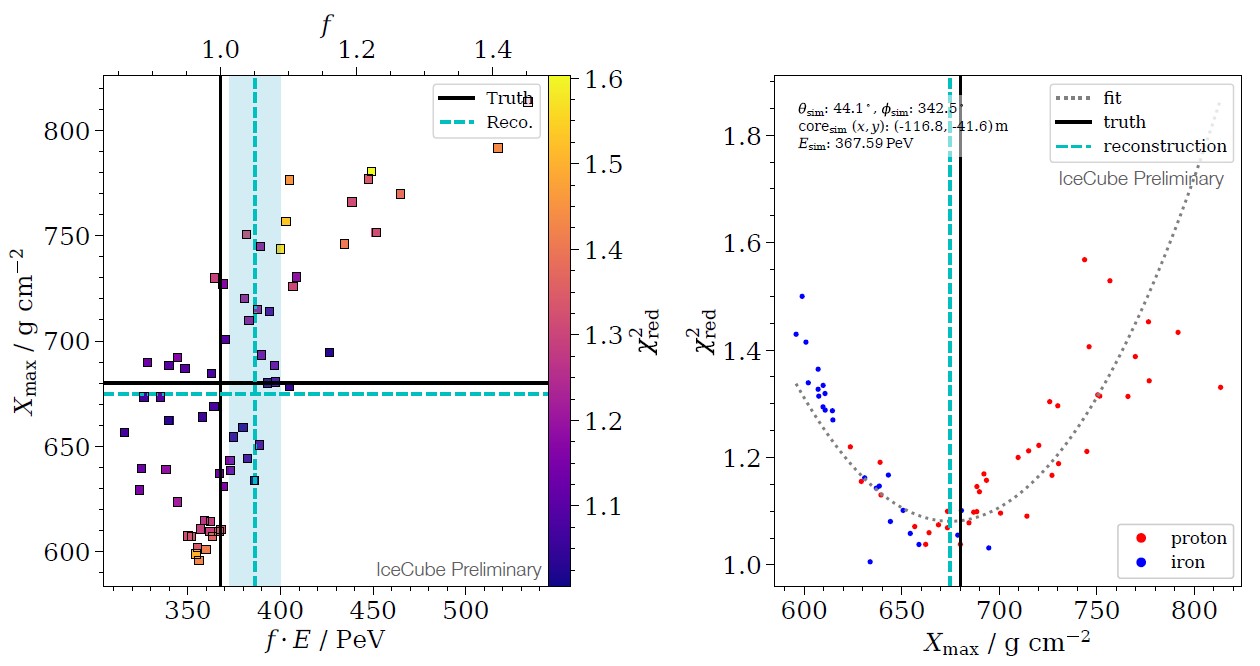}
    \caption{Figure from \refer{Xmaxpaper} shows the minimization applied to mock data. The black solid lines represent the true $X_\mathrm{max}$ and the blue band shows the uncertainty on the scalable fitting parameter. The minimum of the parabola on the right is the reconstructed $X_\mathrm{max}$. }
    \label{fig:Xmaxmockdata}
\end{figure}

The measured detector response along with added measured noise is forward folded to the obtained pulse from CoREAS to obtain mock data. 
The radio data in all measured air showers is cleaned and filtered in the 80 -- 300\,MHz range to find the pulse. The peak of the Hilbert envelope from the waveform was the used measurable in the analysis. A ${\chi^2}$ minimization was performed on the measurable for both the mock and real data with respect to the noiseless simulation. The corresponding $X_\mathrm{max}$ is plotted and compared to the mock data shown in Figure \ref{fig:Xmaxmockdata}. This was the first successful reconstruction method using only 3 antennas in the 80 -- 300\,MHz range.

\section{The Surface Array of IceCube-Gen2} \label{5}
IceCube-Gen2 is the next generation of the IceCube Neutrino Observatory \cite{IceCube-Gen2Whitepaper}. The surface array of Gen2 is expected to build on the design of the IceTop Enhancement. Along with an increase of surface area by a factor of 6, the aperture of the detector is to increase by a factor of around 30, including data from all detectors. This increase in aperture should result in the increase in measured zenith for vetoing atmospheric background for neutrino astronomy, retaining one of the primary motivations behind a surface array. The system is expected to have a global trigger with individual detectors capable of triggering other detectors. More inclined showers can be detected in radio as compared to the future Enhancement due to the possibility of in-ice triggers. With more information from different detectors for a single event, a much more accurate mass estimation of the primary can be performed with IceCube-Gen2. A detailed overview was done in \refer{AlanICRC2023}. 

\section{Summary and a brief Outlook}
We presented a brief overview on all the recent developments in the cosmic-ray detection with the IceTop Enhancement. Preparations for future detector deployments are underway with various instruments being characterized and tested. The use of CNNs to find radio pulses shows much promise and can potentially be extended to other experiments as well. 
The indirect means of detecting air showers adds importance to the parameter $X_\mathrm{max}$ that gives a view into the direct development of the air shower. Radio detection provides the means to estimate $X_\mathrm{max}$ and can be used for mass estimation. The Enhancement establishes the road ahead for the future of the Surface Array for IceCube-Gen2.

\bibliographystyle{ICRC}
\bibliography{refs}


\end{document}